\documentclass[11pt,a4paper]{article}
\usepackage{jinstpub}

\title{Scintillation light detection in the 6-m drift length ProtoDUNE Dual Phase liquid argon TPC}
\author{I. Gil-Botella for the DUNE Collaboration}
\affiliation{CIEMAT, Centro de Investigaciones Energéticas Medioambientales y Tecnológicas,\\ E-28040 Madrid, Spain}
\emailAdd{ines.gil@ciemat.es}

\abstract{The Deep Underground Neutrino Experiment (DUNE) is a leading-edge experiment for long-baseline neutrino oscillation studies, neutrino astrophysics and nucleon decay searches. ProtoDUNE-Dual Phase (DP) is a 6x6x6 m$^3$ liquid argon time-projection-chamber (LArTPC) operated at the CERN Neutrino Platform in 2019-2020 as a prototype of the DUNE Far Detector. In ProtoDUNE-DP, the scintillation and electroluminescence light produced by cosmic muons in the LArTPC is collected by photomultiplier tubes placed up to 7 m away from the ionizing track. In this paper, we present the performance of the ProtoDUNE-DP photon detection system, comparing different wavelength-shifting techniques and the use of xenon-doped LAr as a promising option for future large LArTPCs. The scintillation light production and propagation processes are analyzed and compared to simulations, improving understanding of the liquid argon properties.} 


\proceeding{LIDINE2021 - Light Detection in Noble Elements\\
14-17 September 2021\\
University of California San Diego}

\begin{document}

\maketitle

\section{DUNE and its prototypes at CERN}

The Deep Underground Neutrino Experiment (DUNE)~\cite{DUNE:2020lwj} aims to address key questions in neutrino physics such as measuring the CP violating phase and the neutrino mass ordering with an intense muon neutrino beam produced at Fermilab~\cite{DUNE:2020jqi}. The physics program also addresses non-beam physics such as nucleon decay and Beyond the Standard Model searches~\cite{DUNE:2020fgq} and the detection of astrophysical neutrinos from a core-collapse supernova within the galaxy~\cite{DUNE:2020zfm}. DUNE will consist of a neutrino beam and a near detector placed at Fermilab, and four 17 kt liquid-argon time-projection chambers (LArTPCs) as far detector located underground (4300 m.w.e.~depth) at the Sanford Underground Research Facility (SURF) 1300 km away from Fermilab.

DUNE has been pursuing two LArTPC technologies for the detector modules, single-phase (SP) (liquid only) and dual-phase (DP) (liquid and gas). Two large-scale LArTPC DUNE prototypes were installed and operated at the CERN Neutrino Platform: ProtoDUNE-SP~\cite{DUNE:2020cqd, DUNE:2021hwx} and ProtoDUNE-DP~\cite{DeBonis:2014jlo}. Each of them contains 800 t of LAr making them the largest LArTPCs ever built.

ProtoDUNE-DP was operated from 2019 to 2020 at the CERN Neutrino Platform. The detector has an active volume of 6$\times$6$\times$6 m$^{3}$ corresponding to an active mass of 300\,t. In ProtoDUNE-DP the electric drift field is oriented in the vertical direction, causing the electrons to drift vertically from the cathode at the bottom of the detector towards the anode at the top. The ionization charge is extracted, amplified, and detected in gaseous argon above the liquid surface by the charge readout planes (CRPs), allowing a good signal to noise ratio and a fine spatial resolution. 
The scintillation light signal is collected by a photon detection system (PDS) constructed out of photo-multiplier tubes (PMTs). The PDS goals are to provide a trigger and to determine precisely the event time, with possibility to perform calorimetric measurements and particle identification. Two Cosmic Ray Tagger (CRT) panels 
are placed on opposite sides of the ProtoDUNE-DP cryostat to trigger on muon-tracks passing through both CRTs.

\begin{figure}[ht]
    \centering
    \includegraphics[width=0.55\textwidth]{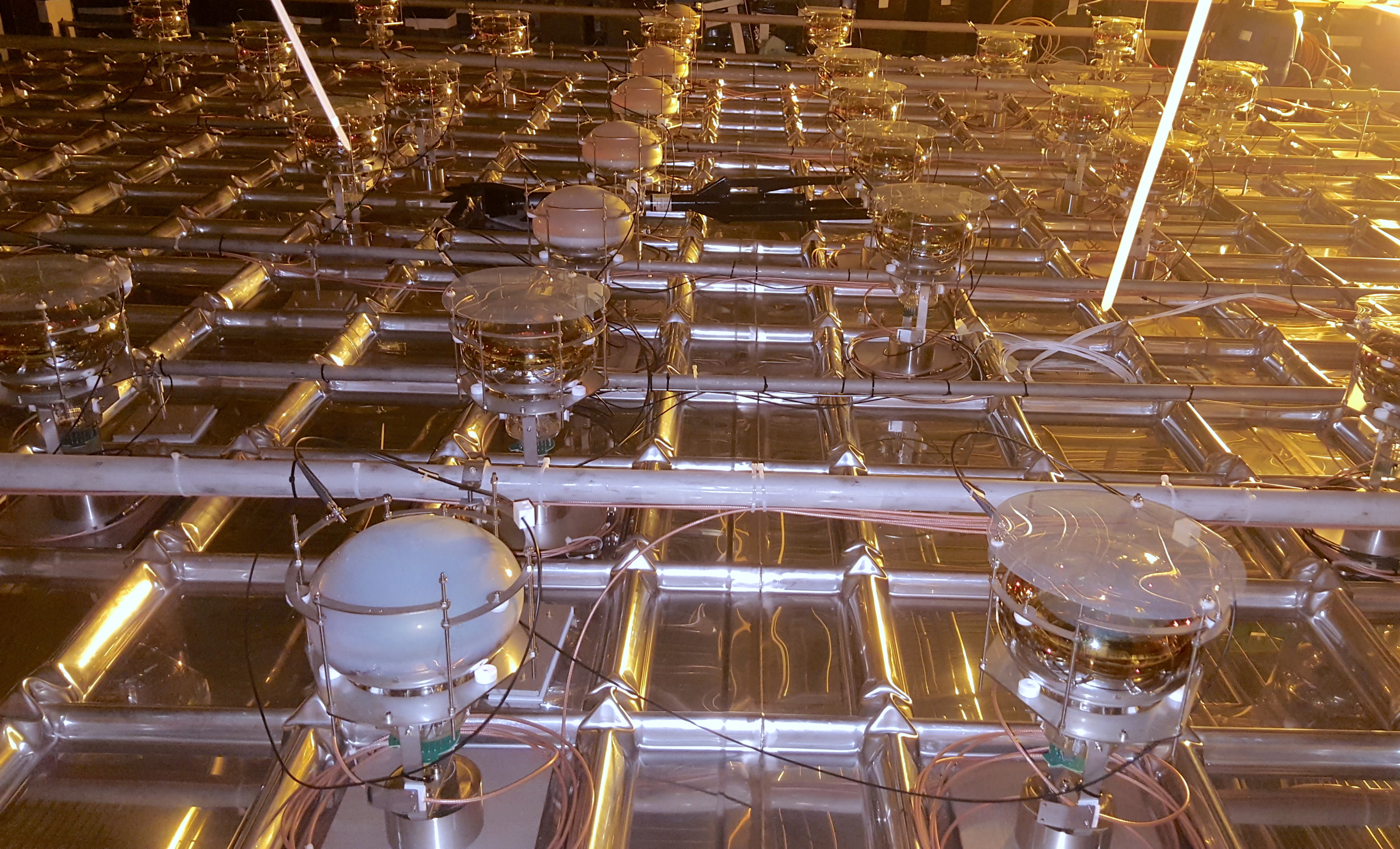}
    \caption{ProtoDUNE-DP PDS formed of 36 PMTs installed at the bottom of the cryostat at CERN. The front left PMT has TPB coating and the front right PMT has PEN sheet.} 
    \label{fig:pds}
\end{figure}

The PDS of ProtoDUNE-DP is formed of 36 8-inch cryogenic R5912-02MOD PMTs from Hamamatsu~\cite{Belver:2018erf, Belver:2019lqm, Belver:2020qmf, Belver:2021drc}, placed at the bottom of the detector below the cathode. As the PMTs are not sensitive to VUV light, a wavelength shifter converts 127-nm photons into visible photons. Two different wavelength shifters were deployed. A sheet of polyethylene naphthalate (PEN) is placed on top of 30 PMTs and the other 6 PMTs have tetraphenyl butadiene (TPB) directly coated on them. Figure~\ref{fig:pds} shows a picture of the PDS installed in ProtoDUNE-DP.

\section{The ProtoDUNE-DP Photon Detection System performance}

ProtoDUNE-DP collected cosmic-ray data from June 2019 until November 2020, operating with pure LAr and Xe-doped LAr (table~\ref{tab:XenonData}). A total of 130.7 million events were acquired with a duration of 675 hours. ProtoDUNE-DP operated fully filled with LAr from August 2019 until May 2020. In June 2020, an intervention on the HV extender was carried out with the aim of fixing the short circuit although the issue was not solved. In July 2020, the detector was re-filled using $\sim$230 ton of Xe-doped LAr from ProtoDUNE-SP contaminated with N$_2$. In August 2020, operations were resumed and two additional N$_2$ injections took place to measure the effect of N$_2$ contamination in the light attenuation length.

\begin{table}[!ht]
\begin{center}
\begin{tabular}{c c c}
\hline
    Situation & [Xe] (ppmm) & [N$_2$] (ppmv) \\
    \hline
    LAr & 0 & 0 \\
    LAr + Xe + N$_{2}$ & 5.8 &  1.7 \\
    1$^{st}$ N$_{2}$ injection & 5.8 & 2.7\\
    2$^{nd}$ N$_{2}$ injection & 5.8 & 4.7 \\
    \hline
\end{tabular}
\caption{Summary of the Xe doping and N$_2$ contamination conditions with the same amount of liquid. The concentrations are expressed in ppmm (parts per million of mass) and ppmv (parts per million volume).}
\label{tab:XenonData}
\end{center}
\end{table}

All 36 PMTs were operational since the beginning of the data taking and the basic performance of the PDS system is validated. A time accuracy among the channels better than 16\,ns is measured. The low noise in the baseline of the signals, with a mean standard deviation of $0.6\pm0.1$\,ADC counts, is remarkable as the baseline presents very small fluctuation and was stable with time. At a gain of $10^7$, the single photo-electron (SPE) amplitude is 7\,ADC counts, implying a signal-to-noise ratio greater than 11 thanks to the small fluctuation of the baseline in the PMT waveforms.



ProtoDUNE-DP uses PMTs either covered with PEN foils or directly coated with TPB
to shift the LAr scintillation photon wavelength towards the visible range.
While TPB is broadly used, PEN is a novel material, never used before in such a large scale experiment and whose efficiency is not well known. As TPB needs complex coating setups, the potential benefit of PEN comes from its simple handling, because PEN foils are flexible plastic sheets easy to fabricate and install. 

The relative photon detection efficiency of the PEN-foil PMTs versus the TPB-coated PMTs, $NPE_\mathrm{PEN}/NPE_\mathrm{TPB}$, is experimentally determined by comparing the amount of light (in PEs) detected by a pair of PEN-TPB PMTs placed symmetrically with respect to the detector and the light source. The average light collected on the PMTs for the selected events is $\sim$200\,PEs on TPB-coated PMTs, and $\sim$50\,PEs on PEN-foil PMTs. An average $NPE_\mathrm{PEN}/NPE_\mathrm{TPB}$ ratio of $0.25\pm0.03$ is obtained. The error is the standard deviation among PMT-pairs, which agrees with the expected error due to the QE variation between PMTs. On average, ProtoDUNE-DP TPB-coated PMTs detect four times more photons than PEN-foil PMTs. Additionally, a simple model is proposed to compute the relative WLS efficiency of both materials taking into account the geometrical differences between both systems. As a result, it is estimated that TPB produces three times more visible photons than PEN, for the same amount of incident VUV photons. 

The scintillation light emission in LAr has a characteristic time dependence. Waveforms are well described by a sum of three exponential functions convoluted with a Gaussian function to represent the detector response. Although the scintillation time profile should in principle have only two components, from the decay to ground state of singlet ($\tau_\mathrm{fast}$) and triplet ($\tau_\mathrm{slow}$) argon excimers, an intermediate component ($\tau_\mathrm{int}$) is added in order to improve the fit as reported also by other LAr experiments~\cite{WA105:2021zin}. To get the scintillation decay times from the PMT waveforms, signals from cosmic muons are selected by triggering on a TPB-coated PMT with a minimum amplitude of 25\,PEs. A $\tau_\mathrm{slow}$ value of $1.46 \pm 0.02$\,$\mu$s is measured, with the error corresponding to the standard deviation among the PMT waveforms. 
The absolute value indicates a high LAr purity at the ppb level. No significant difference is observed in $\tau_\mathrm{slow}$ between PEN and TPB PMTs.

\begin{figure}[ht]
    \centering
    \includegraphics[width=0.43\textwidth]{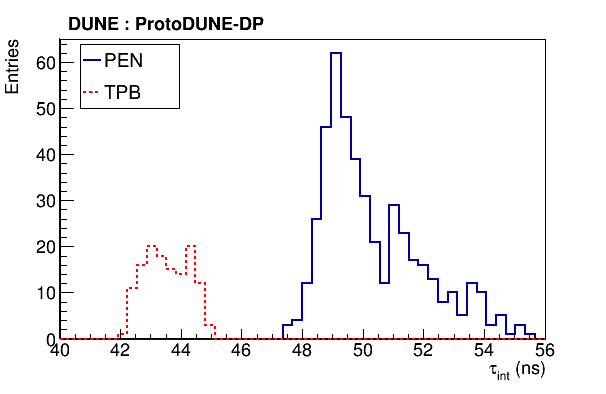}
    \includegraphics[width=0.38\textwidth]{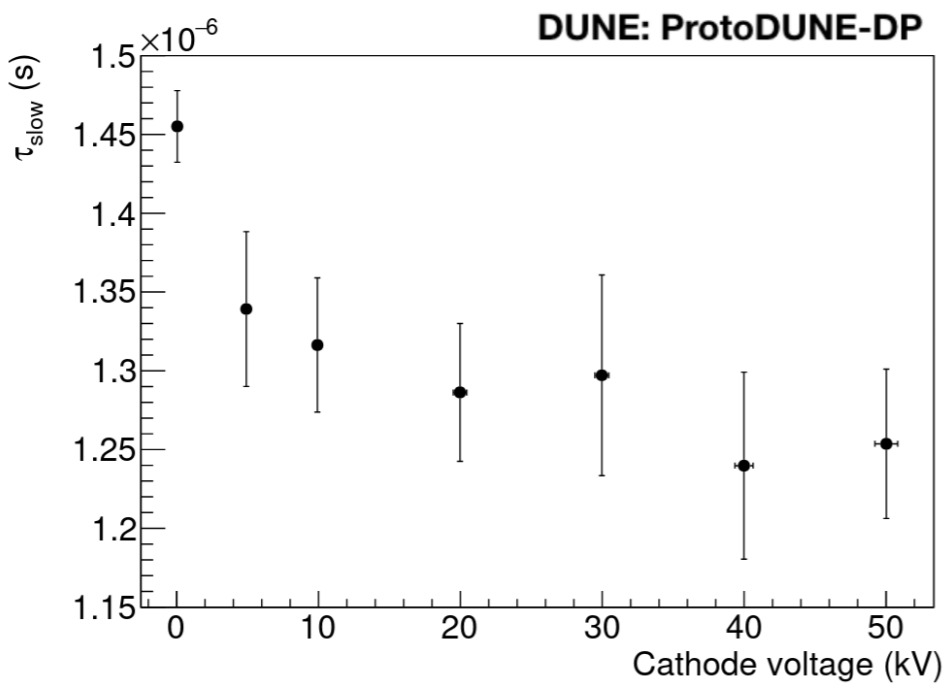}
    \caption{(Left) $\tau_\mathrm{int}$ distribution obtained from the fit to the average waveform for PEN (blue) and TPB PMTs (red). (Right) Evolution of $\tau_\mathrm{slow}$ with the cathode voltage for CRT-trigger events. The results are obtained averaging PEN PMTs and the error bars indicate the standard deviation among them.}
    \label{fig:PENTPB_TauInt}
\end{figure}

Figure~\ref{fig:PENTPB_TauInt} (left) shows the value of $\tau_\mathrm{int}$ for PEN and TPB PMTs obtained from the fit of the average waveforms using data sets for which the purity was already stable. An average value of $50.3 \pm 1.7$\,ns is obtained for the PEN PMTs, and a faster response of $43.6 \pm 0.7$\,ns for the TPB PMTs. The clear difference between the two different WLS points to a delayed emission time by the WLS material, as proposed in~\cite{Segreto:2020qks, Whittington:2014aha}. Figure~\ref{fig:PENTPB_TauInt} (right) shows a decrease of $\tau_\mathrm{slow}$ with the ProtoDUNE-DP drift field, as reported in~\cite{WA105:2021zin}. A model is proposed in~\cite{Segreto:2020qks}, taking into account the quenching of the long-lived triplet states through the self-interaction with other triplet states or through the interaction with molecular Ar$^+_2$ ions.

\section{ProtoDUNE-DP cosmic-muon data and MC simulation}

The study of the scintillation light production, propagation and collection in a LArTPC is performed with data acquired with the CRT-trigger system to profit from the off-line reconstruction of the track trajectory. Analyses are mainly based on the correlation between the scintillation light signal (S1 charge in number of PEs) and the distance from the muon track to the PMT. Muons crossing the CRT panels of ProtoDUNE-DP are simulated using CORSIKA~\cite{CORSIKA}. The photon propagation in LAr, from the production point to the PMT array, is performed with Geant4 in LArSoft~\cite{Snider:2017wjd}.

The size of ProtoDUNE-DP, the longest drift-distance LArTPC ever operated, allows for an unprecedented study of the light propagation. The Rayleigh scattering length (RSL) can impact the amount of light collected. An evaluation of the RSL value is carried out by comparing the measured light signals with the light predicted by the MC simulation testing two lengths (61.0\,cm~\cite{Grace:2015yta} and 99.9\,cm~\cite{Babicz:2020den}) obtained in experimental measurements. Modeling the dependence of the light attenuation from the track-PMT distance with a decaying exponential function allows the measurement of the overall attenuation in data and MC and the evaluation of the agreement for the various simulated configurations. 
Figure~\ref{fig:Raileigh} shows the S1 charge-distance correlation fitted to an exponential and the data-MC ratios for the two simulations. Looking at the distribution shape, the agreement between data and the 99.9-cm MC sample is better than with the 61.0-cm value.

\section{Scintillation light in Xe-doped liquid argon}

The use of Xe-doped LAr is a promising alternative to pure LAr for large-scale LArTPCs, since it mitigates the light suppression due to impurities and it also improves the detection efficiency and uniformity. The effect of the presence of Xe and N$_2$ in the detected light in ProtoDUNE-DP is studied using two types of muon-track signals: First, events triggered with a TPB-coated PMT placed at the center of the detector, for which the light is produced at a close distance from the PMTs, and second, CRT-trigger events, for which the PMT-track distance is in the range of 3$-$5\,m.

\begin{figure}[ht]
    \centering
    \begin{minipage}{0.4\textwidth}
    \centering
    \includegraphics[width=0.9\textwidth]{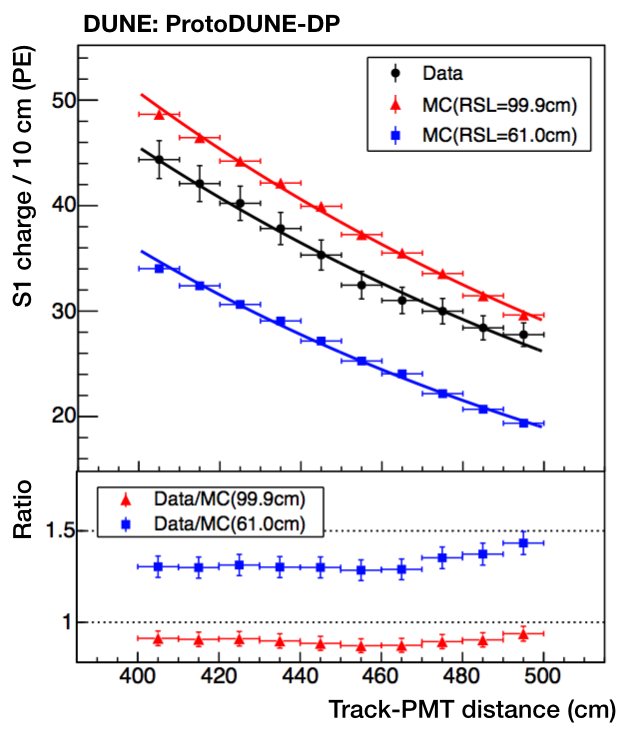}
    \caption{Top panel: Average S1 charge collected by the PEN PMTs as a function of the track-PMT distance. Two MC samples with different RSL values (61.0\,cm and 99.9\,cm) are compared to data. An exponential fit is plotted as a solid line. Bottom panel: Data-MC ratio for the two previous MC samples.} \label{fig:Raileigh}
    \end{minipage}\hspace{0.02\textwidth}
    \begin{minipage}{0.47\textwidth}
    \centering
    \includegraphics[width=0.85\textwidth, height=8.5cm]{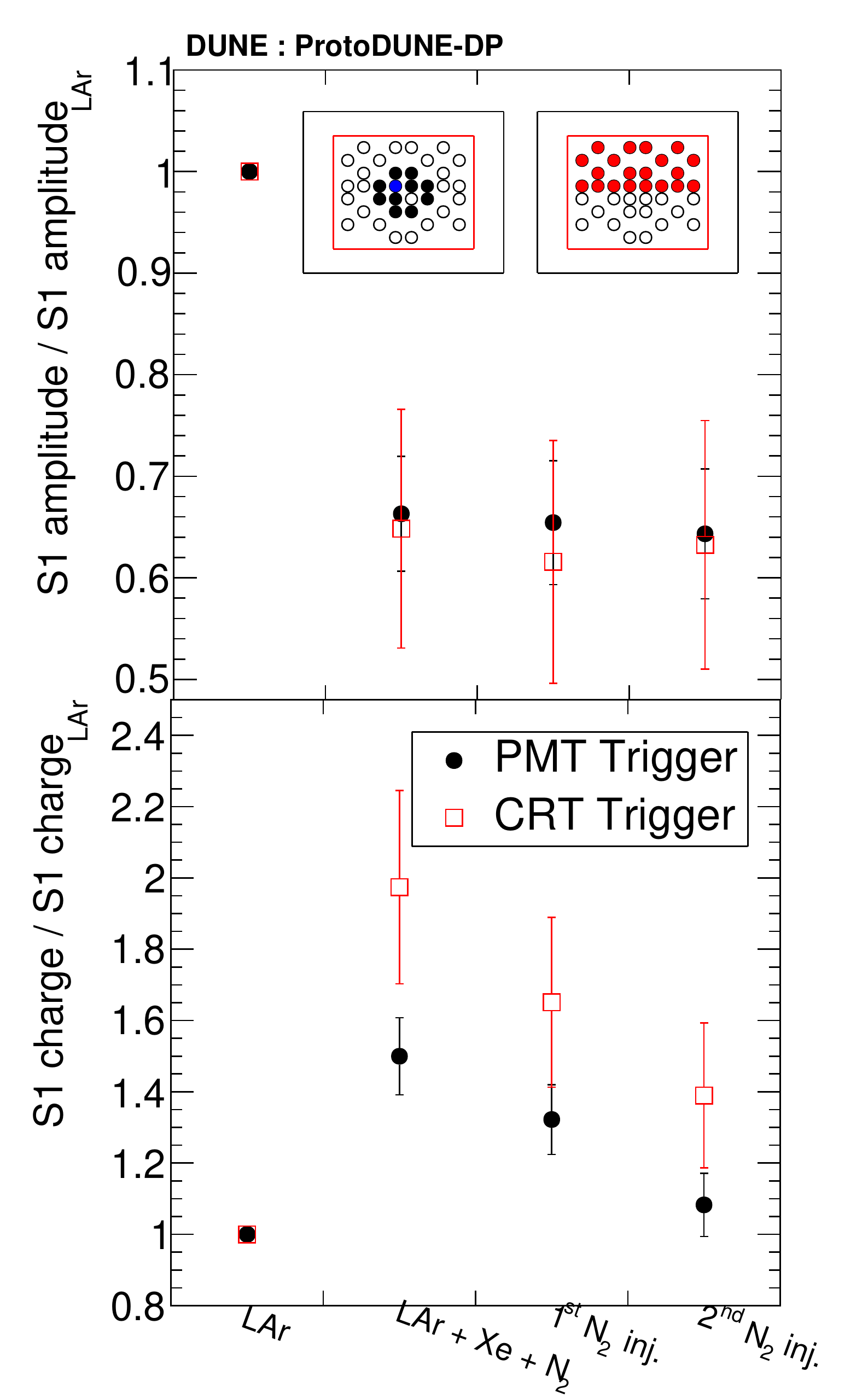}  
    \caption{Average variation of the S1 amplitude and charge in different doping situations relative to pure LAr. PMT (CRT)-trigger data are shown in black (red). Selected PMTs are marked as black (red) circles in the small diagrams.} 
    \label{fig:Xenon_Filled}
    
    \end{minipage} 
\end{figure}

Figure~\ref{fig:Xenon_Filled} shows the average variation of the signal amplitude and S1 charge for different detector conditions. 
The amplitude decreases 35\% when adding 5.8\,ppm of Xe and 1.7\,ppm of N$_{2}$ with respect to pure LAr for both trigger modes, being unaffected by the N$_{2}$ addition. This reduction is due to the absorption of the 127-nm photons by the Xe atoms as reported in~\cite{Neumeier:2015lka}. Whilst a 15\% decrease of the fast component (S1 amplitude) for the CRT-trigger data is expected when adding N$_{2}$ according to~\cite{Jones:2013bca}, we observe the fast component remains constant for the two N$_2$ injections. 
The collected S1 charge increases 100\% for the CRT-trigger data, while only 50\% for the PMT-trigger data. This difference is understood as an improvement of the detection uniformity, since CRT-trigger muons are on average farther away from the PMTs, and the longer RSL of the Xe photons improves their collection at large distances. The decrease of the S1 charge due to the presence of N$_2$ is similar for both triggers (30\%), meaning that there is no dependence of the detected light suppression on the PMT-track distance. This indicates that the reduction is mainly due to the quenching of the Ar excimers by N$_{2}$ rather than photo-absorption.

\section{Conclusions}
The LAr TPC technology at large scale is being demonstrated in the ProtoDUNE Program at the CERN Neutrino Platform for the future DUNE long-baseline experiment. The 6-m drift length ProtoDUNE-DP operated in 2019-2020 at CERN. The photon detection system collected cosmic-ray data for 18 months in stable conditions with all 36 PMTs operative. The good performance of the system has validated the design for use in future long drift distance LArTPCs.
ProtoDUNE-DP used PEN as a wavelength shifter for the first time in a large scale experiment and a comparison with the widely used TPB was carried out. TPB is estimated to be 3 times more efficient than PEN. The size of ProtoDUNE-DP, the longest drift-distance LArTPC ever operated, allows for an unprecedented study of the light propagation. The agreement between data and the 99.9-cm Rayleigh scattering length MC sample is better than for the shorter scattering length.
Finally, ProtoDUNE-DP data has demonstrated the improvement of the light detection efficiency and uniformity in large LArTPCs due to Xe doping. A low doping level of 5.8\,ppm of Xe doubles the collected light at large distances (3$-$5\,m from the PMTs) even in the presence of 1.7\,ppm of N$_{2}$. However, it must be considered that the reduction observed in the light signal amplitude could compromise the performance of a light-based trigger.

\end{document}